# Nexus networks in carbon honeycombs

Yuanping Chen[1]*, Yuee Xie[1], Yan Gao[1], Po-Yao Chang[2],

Shengbai Zhang[3], and David Vanderbilt[2]

[1]*School of Physics and Optoelectronics, Xiangtan University, Xiangtan, 411105, Hunan, China*
[2]*Department of Physics and Astronomy, Rutgers University, Piscataway, New Jersey 08854-8019, USA*
[3]*Department of Physics, Applied Physics, and Astronomy, Rensselaer Polytechnic Institute, Troy, New York, 12180, USA*

**Abstract**

Nexus metals represent a new type of topological material in which nodal lines merge at nexus points. Here, we propose novel networks in nexus systems through intertwining between nexus fermions and additional nodal lines. These nexus networks can be realized in recently synthesized carbon honeycomb materials. In these carbon honeycombs, we demonstrate a phase transition between a nexus network and a system with triply-degenerate points and additional nodal lines. The Landau level spectra show unusual magnetic transport properties in the nexus networks. Our results pave the way toward realizations of new topological materials with novel transport properties beyond standard Weyl/Dirac semimetals.

Corresponding author: chenyp@xtu.edu.cn

**Introduction**

The recent discovery of topological metals (TMs) and semimetals offers new opportunities for observing elementary excitations in the condensed matter world that are not found in particle physics[1-7]. For examples, zero-dimensional (0D) nodal points with multiple degeneracy can be stabilized by crystalline symmetries[8-17], and one-dimensional (1D) objects such as Dirac and Weyl nodal lines can also exist in crystals[18-25]. These 0D and 1D gapless excitations have no analogue in elementary particle physics.

One interesting scenario for gapless excitation is the combined structure of nodal points and lines. In the presence of three mirror planes related by a three-fold rotation and an additional mirror plane perpendicular to the rotation axis, two triple points (TP) connected by a nodal line can be stabilized along the rotation axis[26-30]. By breaking the perpendicular mirror symmetry, this nodal line can split into four topological nodal lines, three of which reside on three mirrors planes while the last one remains on the rotation axis, as shown in Fig. 1(a). In this case, the TPs are denoted as nexus points (NPs), which can be seen as merging points of multiple nodal lines[31-35]. This transition has been predicted in several candidate materials[32-35]. New topological phenomena and transport properties created by nexus fermions have been proposed, such as unusual topological surface states[32], transport anomalies, and topological Lifshitz transitions[35].

Here we propose that the NP phase can further evolve into novel network structures. These networks originate from the original NP phase interacting with additional nodal lines protected by the mirror symmetry, as shown by the orange lines in Fig. 1(b). Because crossings between topological nodal lines are not allowed in the absence of any special symmetry constraint, anticrossing nodal lines are formed as shown in Fig. 1(c). As a result, nodal lines reconnect and the NPs become the intersection points of anticrossing nodal lines, as shown in Fig. 1(d). Considering that the system has three mirror planes, the NPs and the connecting nodal lines form a three-dimensional (3D) network in the full momentum space. We refer to this as a nexus network.

We propose that these nexus networks can be realized in carbon honeycombs (CHCs). CHCs have recently been synthesized, although further refinement of the structure appears highly desirable[36]. We have identified two types of topological networks that can form in the carbon honeycombs. The first type is a nexus network in which the NPs are connected by nodal lines through two kinds of connectivities: a standard connectivity and a winding connectivity (we will define these later). The second type consists of TPs with additional nodal lines (ANLs), named as TP-ANL. These two types of networks correspond to different crystal symmetries. Therefore, the transition between these two networks can be induced by changing the corresponding crystal symmetries. In addition, we show that the Landau level (LL) spectrum can reveal the exotic magnetic transport properties of these two networks.

**Atomic structure of CHC and Computational method**

An atomic structure for the CHC is shown in Fig. 2(a), where carbon atoms form a 3D honeycomb structure. The atoms can be classified according to their characteristic orbital hybridization: $sp^2$ for carbon C1 and $sp^3$ for carbon C2. The C1 atoms can be further subdivided into two subgroups, as shown by the green and blue atoms, forming zigzag chains lying in mirror planes that intersect along the 3-fold axis [see Fig. 2(b)]. One can increase the width of the chains, corresponding to $n = 1$ in Fig. 2(a), to form wider CHC-$n$ nanoribbons with $n > 1$. The primitive cell of CHC-1 is shown in Fig. 2(b). The C2 atoms, on the other hand, form $sp^3$-bonded carbon dimers. In the absence of the dimerization, however, the honeycomb contains only $sp^2$-bonded carbon, as shown in Fig. 2(c), which will be termed CHC-1′ to distinguish it from CHC-1.

According to a space-group analysis, CHC-$n$ can be divided into two types. When $n$ is odd, the space group is P$\bar{3}m1$ with a 3-fold rotational symmetry along $z$ and three mirror planes parallel to $z$. When $n$ is even, the space group is P$6_3$/mmc with a 6-fold screw rotational symmetry along $z$, a mirror plane $M_z$ normal to $z$, and three mirror and three glide planes parallel to $z$. Note that CHC-1′ also has the space group P$6_3$/mmc. Detailed structural parameters for the structures can be found in Table S1 in the supplementary information (SI) [37].

Our first-principles calculations were based on density functional theory (DFT) within the PBE approximation for the exchange-correlation energy [38]. The core-valence interactions were described by the projector augmented-wave (PAW) potentials, as implemented in the VASP code [39-41]. Plane waves with a kinetic energy cutoff of 600 eV were used as the basis set. The atomic positions were optimized via the conjugate gradient method, in which the energy convergence criterion between two consecutive steps was set at $10^{-6}$ eV. The maximum allowed force on the atoms is $10^{-3}$ eV/Å. A 9×9×15 k-point mesh was used for the BZ integration of CHC-1 (or 1×1×2 supercell of CHC-1′). The quadratic crossing leads to a trivial

**Band structure and topological networks**

Figure 3(a) shows the band structure of CHC-1 along $k_z$ (Γ − A), indicating that the carbon structure is metallic. There is a point α (E = 0.50 eV; $k_z = 0.07$ π/$c$ and $c = 4.83$ Å is the lattice constant along $z$ axis) at which the black and green bands cross. Due to the negligibly small spin-orbit coupling strength in carbon, spin is neglected and the electrons are treated as spinless. In this context, the green band is doubly degenerate while the black band is non-degenerate. Therefore, the point α is a TP where three bands $m$-1, $m$ and $m$+1 cross ($m$ = 33 here). Another TP is located at $k_z$ = - 0.07 π/$c$ because of the structural inversion symmetry.

In a $k_x$-$k_y$ plane containing one TP, the TP appears as a crossing point of a Weyl cone and a flat energy band; however, in a $k_x$-$k_z$ (or $k_y$-$k_z$) plane containing both TPs, the two TPs are connected by a nodal line which is intersected by a cone and a tilted energy surface along $k_z$ [see Fig. S1 in SI [37]]. The nodal line corresponds to the green band in Fig. 3(a), in which the solid and dotted parts represent degeneracy between bands ($m$-1,$m$) and ($m$,$m$+1) respectively. An examination clearly indicates that the TPs are also connected by other nodal lines on the three mirror planes $k_y = 0, \pm\sqrt{3}k_x$. Figure 3(d) exhibits the nodal lines for bands ($m$,$m$+1) on the mirror plane $k_y$ = 0, while Fig. 3(e) exhibits those for bands ($m$-1,$m$). The two blue dots correspond to the TPs. As can be seen from Fig. 3(d), the two TPs are connected by a straight line and a curved line both of which pass through the high-symmetry point A. The straight line corresponds to the dashed green line in Fig. 3(a). Because the structure has a 3-fold rotation axis along $z$, the two other mirror planes $k_y = \pm\sqrt{3}k_x$ have identical nodal lines. The dotted lines in Fig. 3(d) show the nodal lines for bands ($m$,$m$+1) in the first Brillouin zone (BZ). It illustrates that this is a standard nexus phase like that in Fig. 1(a), and thus the TPs are in fact NPs connected by four nodal lines. Each mirror plane has a curved nodal line, while the three mirror planes share a straight line. We refer to this form of connection between the two NPs as a standard connectivity.

As seen in Fig. 3(e), there are four nodal lines connecting the two NPs, similar to what was shown in Fig. 1(d). A straight line, corresponding to the solid green line in Fig. 3(a), links the two NPs through Γ along $k_z$. Considering the periodicity of the BZ, the lines off the $k_z$ axis are actually linked to each other through points β and β′, that is, the two NPs are connected by a curved line that starts from one TP and then passes through β, Γ, and β′ before arriving at the other TP, winding twice around the Brillouin torus in the process. We refer to this type of connection between the two NPs as a "winding connectivity". The solid lines in Fig. 3(b) show the winding connectivity in the first BZ. Thus, Fig. 3(b) summarizes the structure of our novel 3D nexus network.

By comparing the standard connectivity and winding connectivity, one can find that the nodal lines in the former remain close to the $k_z$ axis, while those in the latter extend through the full BZ. Therefore, the nexus network may be easier to observe experimentally than the standard NP phase, because it extends the regions in which the topological elements can be found.

When the dimerization is removed, the CHC-1 structure evolves into CHC-1′, whose primitive cell is half that of CHC-1 as shown in Fig. 2(c). For comparison, the band structure for a $1\times1\times2$ supercell of CHC-1′ is calculated. The results indicate that it is a TP-ANL metal, where TPs and ANLs coexist in momentum space. Figure 3(f) shows the nodal lines for bands (*m*-1,*m*) on the $k_y = 0$ mirror plane. Two TPs are located at $k_z = \pm 0.07\ \pi/2c'$ ($c'$ is the lattice constant of CHC-1′), and they are connected by a straight nodal line. In addition, there are two nodal lines along $k_x$. Because the structure has a 6-fold screw symmetry, the other two mirror planes $k_y = \pm\sqrt{3}k_x$ and three glide planes $k_x = 0, \pm\sqrt{3}k_y$ have the same nodal line distributions as for the $k_y = 0$ plane. Figure 3(c) presents the whole topological network in the BZ. The ANLs cross at some points on the $k_z$ axis other than at the TPs.

**Topological phases based on tight-binding model**

We analyze the topological networks [see Figs. 3(b) and 3(c)] from a tight-binding model, which is designed to mimic the DFT band structure near the Fermi level. The bands of CHC-1 near the Fermi level are mainly contributed by π orbitals on the $sp^2$-hybridized C1 atoms [see Fig. S1(b) in SI [37]]. Therefore, it is convenient to describe the carbon honeycomb by a tight-binding model,

$$H = \sum_{<i,j>} \sum_{\mu} t_{ij} e^{-i\bm{k}\cdot \bm{d}_{ij}^{\mu}}, \qquad (1)$$

where i, j ∈ {1,2, ... ,12} label the twelve $C_1$ atoms, $d_{ij}^{\mu}$ is the displacement vector directed from atoms j to i, $t_{ij}$ is the hopping energy between atoms i and j, and μ runs over all lattice sites under translation. For $t_{ij}$, we include hoppings $t_0$ between atoms in the zigzag chains and hoppings $t_1$ ~ $t_5$ between the zigzag chains [see Figs. 2(a-b)].

To reveal the effect of structural symmetry on the electronic properties, we first consider the 1×1×2 supercell of CHC-1′. It has space group P6$_3$/mmc, which includes a 6-fold screw rotational symmetry along z, a mirror plane $M_z$ normal to z, and three mirror planes and three glide planes parallel to z. If we omit its C2 atoms, its electronic properties can also be described by Eq. (1). Due to the high symmetry, the values of $t_0$ ~ $t_5$ are simpler. Two sub-cases are considered according to $t_2$. When $t_2$ = 0, the tight-binding model represents a simple TP phase, in which two TPs are connected by a nodal line along $k_z$. Figure 4(a) shows the phase on one mirror plane $k_y$ = 0. A further calculation indicates that each point on the nodal line is a crossing point of quadratic bands on the $k_x$-$k_y$ plane [the bands along $k_x$ are shown in Fig. 4(a1)]. The quadratic crossing leads to a trivial Berry phase of 2π, which was computed for a closed loop around the crossing point. That is, the two TPs are connected by a $Z_2$-trivial line. When $t_2$≠0, a TP-ANL phase like that of Fig. 3(c) is generated. In each mirror/glide plane, as shown in Fig. 4(b), two (orange) ANLs are now presented. Figure 4(b1) illustrates that each point on these ANLs is a crossing point of linear bands, leading to a nontrivial Berry phase of π. The crossing between the trivial green line and nontrivial orange lines is allowed because of protection by the structural symmetry. Following the evolution from Figs. 4(a) to 4(b), one can find that the parameter $t_2$ induces the ANLs on the mirror and glide planes.

We then consider the structure CHC-1. The dimers decrease the structural symmetry, which leads to the elimination of the three glide planes and the horizontal mirror plane $M_z$ in CHC-1′. Atomic bond lengths of the structure become complicated, and thus the values of $t_0 \sim t_5$ become complicated. Two other sub-cases are now considered. When $t_2 = 0$, the tight-binding model produces a standard NP phase. Figure 4(c) shows the nodal line structure on one mirror plane $k_y = 0$. Comparing Figs. 4(a) and 4(c), one can see that the reduction of the symmetry causes the trivial nodal line in the TP phase to split into two nontrivial lines in the NP phase. This splitting originates from the fact that the quadratic points in Fig. 4(a1) split to two nodal points in Fig. 4(c1).

When $t_2 \neq 0$, the novel nexus network in Fig. 3(d) is reproduced. Figure 4(d) shows the network on one mirror plane $k_y = 0$. The two NPs are connected by standard connectivity for bands $(m,m+1)$, while they are connected by winding connectivity for bands $(m-1,m)$. The formation of the nexus network originates from the interactions between the standard NP phase and the ANLs. In the case of $t_2 \neq 0$, the NP phase in Fig. 4(c) will overlap with the ANLs, which seems like Fig. 1(b). A crossing of two nontrivial lines is not generally allowed if there are no special symmetry to protect the crossing [42]. The nodal lines at the crossing point will split into two anti-crossing lines, as shown in Fig. 1(c). As a consequence, the nodal lines in Fig. 1(b) evolve to adopt the structure shown in Fig. 1(d) or 4(d). [Detailed information about the tight-binding model is presented in the SI.]

**k · p model and Landau levels**

Constrained by the symmetry groups and the time reversal symmetry for spinless systems, one obtains a k · p model around the Γ point:

$$H(\boldsymbol{k}) = \begin{bmatrix} A_1 k_\parallel^2 + B_1 \cos k_z + C_1 & \alpha k_+ \sin k_z + \beta k_-^2 & Dk_- \\ \alpha k_- \sin k_z + \beta k_+^2 & A_1 k_\parallel^2 + B_1 \cos k_z + C_1 & -Dk_+ \\ Dk_+ & -Dk_- & A_2 k_\parallel^2 + B_2 \cos k_z + C_2 \end{bmatrix},$$

(2)

where $k_\pm = k_x \pm i k_y$, $k_\parallel^2 = k_x^2 + k_y^2$, and $A_{1,2}$, $B_{1,2}$, $C_{1,2}$, D, $\alpha$, $\beta$ are real constants. When $\alpha = 0$, this model describes the topological phases in CHC-1′ with 6-fold rotational symmetry. When $\alpha \neq 0$, it describes those in CHC-1 because the $\alpha k_+ \sin k_z$ term decreases the symmetry to 3-fold rotational symmetry. For the phases in CHC-1′, the k · p model with $\beta = 0$ and $\beta \neq 0$ corresponds the TP phase in Fig. 4(a) and TP-ANL phase in Fig. 4(b), respectively. Therefore, the effect of the $\beta k_\pm^2$ term is to generate ANLs on the mirror/glide planes. For the phases in CHC-1, Eq. (2) with $\beta = 0$ and $\beta \neq 0$ describes a standard NP phase as in Fig. 4(c) and a nexus network as in Fig. 4(d), respectively. Therefore, the effect of the $\alpha k_\pm \sin k_z$ term is to split the trivial line in the TP phase. This further shows that the nexus network ($\alpha \neq 0$, $\beta \neq 0$) results from the interactions between the standard NP phase and the ANLs.

The reason to develop a k · p model here is not only because such a generic model can reproduce the nexus network of CHC-1 [see Fig. S3 in SI [37]], but more importantly it depicts a more complete picture of the often-complicated nexus networks, which may not exist in the CHC's but may exist in other materials. For example, if $\beta$ in Eq. (2) is replaced by $\beta + \gamma \cos(\delta k_z)$, the different $\gamma$ and $\delta$ will generate rich TP-ANL phases and nexus networks [see Fig. S4 in SI [37]]. Figure 5(a) presents a TP-ANL that includes four ANLs in each mirror/glide plane, while Fig. 5(b) presents the corresponding nexus network that results when the TPs transition to NPs.

Based on Eq. (2), we calculate the LLs for the TP-ANL and nexus networks. As shown in Figs. 4(c-f), besides the gapped LLs away from the Fermi level, there are gapless chiral LLs cross the Fermi level. As mentioned above, the straight nodal line connecting TPs or NPs along $k_z$ are intersected by a Weyl cone and a tilted energy surface. The gapped LLs are related to the cone, while the gapless chiral LLs originate from the tilted energy surface. Different behaviors of the chiral LLs come from different shapes of energy surfaces. For example, at the mirror plane $k_y = 0$, the eigenvalue of the tilted surface is

$$E_1 = (A_1 + \beta)k_x^2 + \alpha k_x \sin k_z + B_1 \cos k_z + C_1. \tag{3}$$

For the TP-ANL phases ($\alpha = 0$), when $A_1 = -\beta$, $E_1$ is a constant for a certain $k_z$. This means that the energy surface is completely flat along the direction normal to $k_z$. In this case, the chiral LLs collapse to one LL, as shown in Fig. 5(c). For the nexus network ($\alpha \neq 0$), if $A_1 = -\beta$, the chiral LLs become degenerate when $k_z = 0, \pm\pi$, while remaining split at other $k_z$, as shown in Fig. 5(d). Because of the structural 3-fold symmetry, the splitting LLs have 3-fold degeneracy.

For the complicated systems in Figs. 5(a) and 5(b), the chiral LLs become exotic. Figures 5(e) and 5(f) show their LLs, respectively. In Fig. 5(e), the chiral LLs form a "standing wave" shape, with the wave nodes located at $k_z = \pm\frac{n}{8}\pi$ for $n$ odd. This is because, when $\gamma \cos(\delta k_z) = 0$, the equation for the energy surface in this case becomes identical to Eq. (3). This constraint leads to $k_z = \pm\frac{n}{8}\pi$ for the parameters given in Figs. 5(a) and 5(b) ($\delta = 4$). Therefore, degenerate LLs appear at $k_z = \pm\frac{n}{8}\pi$ for the TP-ANL phase ($\alpha = 0$). For the nexus phase in Fig. 5(b), the chiral LLs split because $\alpha \neq 0$, as shown in Fig. 5(f). These exotic LLs not only exhibit "fingerprints" for the different topological phases, but also provide novel magnetic transport properties.

**Discussion**

The space group of CHC-$n$ with even $n$ is P6$_3$/mmc, similar to that of CHC-1′. Therefore, TP-ANL phases exist in CHC-$n$ ($n$=2, 4, …). Moreover, with the increase of $n$, there exist four-fold degenerate nodal points located between the TPs, which result in ANLs [see Fig. S5 in SI [37]]. The structures CHC-$n$ with odd $n$ belong to the space group P$\bar{3}m1$, and all of them have nexus networks like CHC-1. Although the experiment in Ref. [36] reported that both periodic and random CHC structures have been obtained, these topological phases maybe are not easy to be observed in the recent samples because the crystal quality needs further improvement. We hope more experiments can be carried out to validate and extend our findings.

The topological classification of the nexus networks is different from that of standard NP phase. The nodal lines in the latter can be continuously contracted to a straight line. Those in the nexus networks, however, wind around the entire BZ torus and thus are not contractible. Mathematically, the BZ is topologically equivalent to a three-dimensional torus T$^3$. The connecting lines on T$^3$ can be classified under its fundamental homotropy group Z$^3$[43], labeled by three integers, each indicating the number of times the loop winds around one of the three directions $k_a$, $k_b$ and $k_c$. In this sense, the standard TP phase in Fig. 4(a) and standard NP phase in Fig. 4(c) belong to the trivial class with Z$^3$ = (0,0,0). However, the TP-ANL phases in Figs. 4(b) and 5(a) can be characterized as Z$^3$ = (1,0,0), and the nexus networks in Figs. 4(d) and 5(b) can be characterized as Z$^3$ = (2,0,0).

Nexus networks originate from the interactions between the standard NP phase and additional nodal lines. This opens a door to propose and search for complex novel topological phases based on simpler topological elements. For example, the interactions between nodal lines/rings may lead to 3D nodal line networks [44], and the coexistence of multiple-fold (3-, 4-, 6- and 8-fold) degenerate points and nodal lines may generate more complicated new 3D networks. The new topological networks not only bring emergent particles and novel concepts, but also create new topological phenomena and transport behaviors.

**Acknowledgments:**

We thank Damien West for useful discussions. YPC and YEX were supported by the National Natural Science Foundation of China (Nos. 11474243 and 51376005). PYC was supported by the Rutgers Center for Materials Theory. SBZ was supported by the US Department of Energy (DOE) under Grant No. DESC0002623. DV was supported by the NSF DMR-1408838.

# References

[1] S.-Y. Xu, et al. Discovery of a Weyl fermion semimetal and topological Fermi arcs. *Science* **349**, 613 (2015).

[2] B. Q. Lv, et al. Experimental discovery of Weyl semimetal TaAs. *Phys. Review X* **5**, 031013 (2015).

[3] S. Y. Xu, et al. Observation of Fermi arc surface states in a topological metal. *Science* **347**, 294 (2015).

[4] L. Lu, et al. Experimental observation of Weyl points. *Science* **349**, 622 (2015).

[5] B. Q. Lv, et al. Observation of three-component fermions in the topological semimetal molybdenum phosphide. *Nature* **546**, 627 (2017).

[6] B. Yan and C. Felser. Topological Materials: Weyl Semimetals. *Annu. Rev. Conden. Ma. P.* **8**, 337 (2017).

[7] N. P. Armitage, E. J. Mele, and Ashvin Vishwanath. Weyl and Dirac semimetals in three-dimensional solids. *Rev. Mod. Phys.* **90**, 015001 (2018).

[8] B. Bradlyn, et al. Beyond Dirac and Weyl fermions: unconventional quasiparticles in conventional crystals. *Science* **353**, aaf5037 (2016).

[9] Z. Liu, et al. Discovery of a three-dimensional topological Dirac semimetal, Na3Bi. *Science* **343**, 864 (2014).

[10] A. A. Soluyanov, et al. Type-II Weyl semimetals. *Nature* **527**, 495 (2015).

[11] S. Borisenko, et al. Experimental realization of a three-dimensional Dirac semimetal. *Phys. Rev. Lett.* **113**, 027603 (2014).

[12] Y. Chen, et al. Nanostructured carbon allotropes with Weyl-like loops and points. *Nano Lett.* **15**, 6974 (2015).

[13] R. Batabyal, et al. Visualizing weakly bound surface Fermi arcs and their correspondence to bulk Weyl fermions. *Sci. Adv.* **2**, e1600709 (2016).

[14] X. Wan, et al. Topological semimetal and Fermi-arc surface states in the electronic structure of pyrochlore iridates. *Phys. Rev. B* **83**, 205101 (2011).

[15] H. Weng, et al. Weyl semimetal phase in noncentrosymmetric transition-metal monophosphides. *Phys. Rev. X* **5**, 011029 (2015).

[16] Z. Wang, et al. Dirac semimetal and topological phase transitions in $A_3$Bi (A=Na, K, Rb). *Phys. Rev. B* **85**, 195320 (2012).

[17] S. M. Young, et al. Dirac semimetal in three dimensions. *Phys. Rev. Lett.* **108**, 140405 (2012).

[18] A. A. Burkov, M.D. Hook, L. Balents. Topological nodal semimetals. *Phys. Rev. B* **84**, 235126 (2011).

[19] C. Fang, et al. Topological nodal line semimetals with and without spin-orbital coupling. *Phys. Rev. B* **92**, 081201(R) (2015).

[20] H. Huang, et al. Topological nodal-line semimetals in alkaline-earth stannides, germanides, and silicides. *Phys. Rev. B* **93**, 201114(R) (2016).

[21] Y. Kim, et al. Dirac line nodes in inversion-symmetric crystals. *Phys. Rev. Lett.* **115**, 036806 (2015).

[22] S. Murakami, et al. Emergence of topological semimetals in gap closing in semiconductors without inversion symmetry. *Sci. Adv.* **3**, e1602680 (2017).

[23] T. Bzdusek, et al. Nodal-chain metals. *Nature* **538**, 75 (2016).

[24] Z. Yan, et al. Nodal-link semimetals. *Phys. Rev. B* **96**, 041103 (2017).

[25] Y. Gao, et al. A class of topological nodal rings and its realization in carbon networks. Preprint at https://arxiv.org/abs/1707.04576 (2017).

[26] C. Zhong, et al. Three-dimensional Pentagon Carbon with a genesis of emergent fermions. *Nature Comm.* **8**, 15641 (2017).

[27] L. Lu, et al. Weyl points and line nodes in gyroid photonic crystals. *Nature Photon.* **7**, 294 (2013).

[28] H. Weng, C. Fang, Z. Fang, X. Dai. Topological semimetals with triply degenerate nodal points in θ-phase tantalum nitride. *Phys. Rev. B* **93**, 241202 (2016).

[29] H.Yang, et al. Prediction of triple point fermions in simple half-Heusler topological insulators. *Phys. Rev. Lett.* **119**, 136401 (2017).

[30] C. Shekhar et al. Extremely high conductivity observed in the unconventional triple point fermion material MoP. https://arxiv.org/abs/1703.03736 (2017).

[31] Z. Yan, et al. Nodal-link semimetals. *Phys. Rev. B* **96**, 041103 (2017).

[32] T. Hyart, T. T. Heikkila. Momentum-space structure of surface states in a topological semimetal with a nexus point of Dirac lines. *Phys. Rev. B* **93**, 235147 (2016).

[33] T. T. Heikkila, G. E. Volovik. Nexus and Dirac lines in topological materials. *New J. Phys.* **17**, 093019 (2015).

[34] G. Chang, et al. New fermions on the line in topological symmorphic metals. *Sci. Rep.* **7**, 1688 (2017).

[35] Z. Zhu, G. W. Winkler, Q. Wu, J. Li, A. A. Soluyanov. Triple point topological metals. *Phys. Rev. X* **6**, 031003 (2016).

[36] N. V. Krainyukova and E. N. Zubarev. Carbon Honeycomb High Capacity Storage for Gaseous and Liquid Species. *Phys. Rev. Lett.* **116**, 055501 (2016).

[37] See Supplemental Material at [URL will be inserted by publisher] for describing tight-binding model and k·p model for CHC-1, the parameters for the two models, DFT band structures of CHC-1 and CHC-2, tight-binding band structure of CHC-1, topological phases induced by the k·p model, structural parameters of CHC-1, CHC-2 and CHC-1'.

[38] J. P. Perdew, et al. Atoms, molecules, solids, and surfaces: applications of the generalized gradient approximation for exchange and correlation. *Phys. Rev. B* **46**, 6671 (1992).

[39] G. Kresse, J. Furthmuller. Efficiency of ab-initio total energy calculations for metals and semiconductors using a plane-wave basis set. *Comput. Mater. Sci.* **6**, 15 (1996).

[40] G. Kresse, J. Hafner, Ab initio molecular dynamics for liquid metals. *Phys. Rev. B* **47**, 558 (1993).

[41] P. E. Blöchl, Projector augmented-wave method. *Phys. Rev. B* **50**, 17953 (1994).

[42] For example, in the condition of time reversal and inversion symmetry, we consider a Hamiltonian:

$$H(k) = f_0(\boldsymbol{k})\sigma_0 + f_1(\boldsymbol{k})\sigma_1 + f_3(\boldsymbol{k})\sigma_3$$

where k = ($k_x$, $k_y$, $k_z$), $\sigma_0$, $\sigma_1$ and $\sigma_3$ are identity, first and third Pauli matrices. The restrictions $f_1(\boldsymbol{k}) = 0$ and $f_3(\boldsymbol{k}) = 0$ apply two constrictions so the touchings are of dimension 1. Many choices for $f_1(\boldsymbol{k})$ and $f_3(\boldsymbol{k})$ are possible to generate two nodal lines. However, to allow the crossing of two lines, we must find a point $k_0$ satisfying $f_1(\boldsymbol{k}) = 0$, $f_3(\boldsymbol{k}) = 0$ and $\nabla f_1(\boldsymbol{k}) = c f_3(\boldsymbol{k})$ (c is a constant).

[43] S. Li, et al. Type-II nodal loops: Theory and material realization. *Phys. Rev. B* 96, 081106(R) (2017).

[44] J. Cai et al. Three-dimensional Hopf chains evolved from triple points. (Submitted)

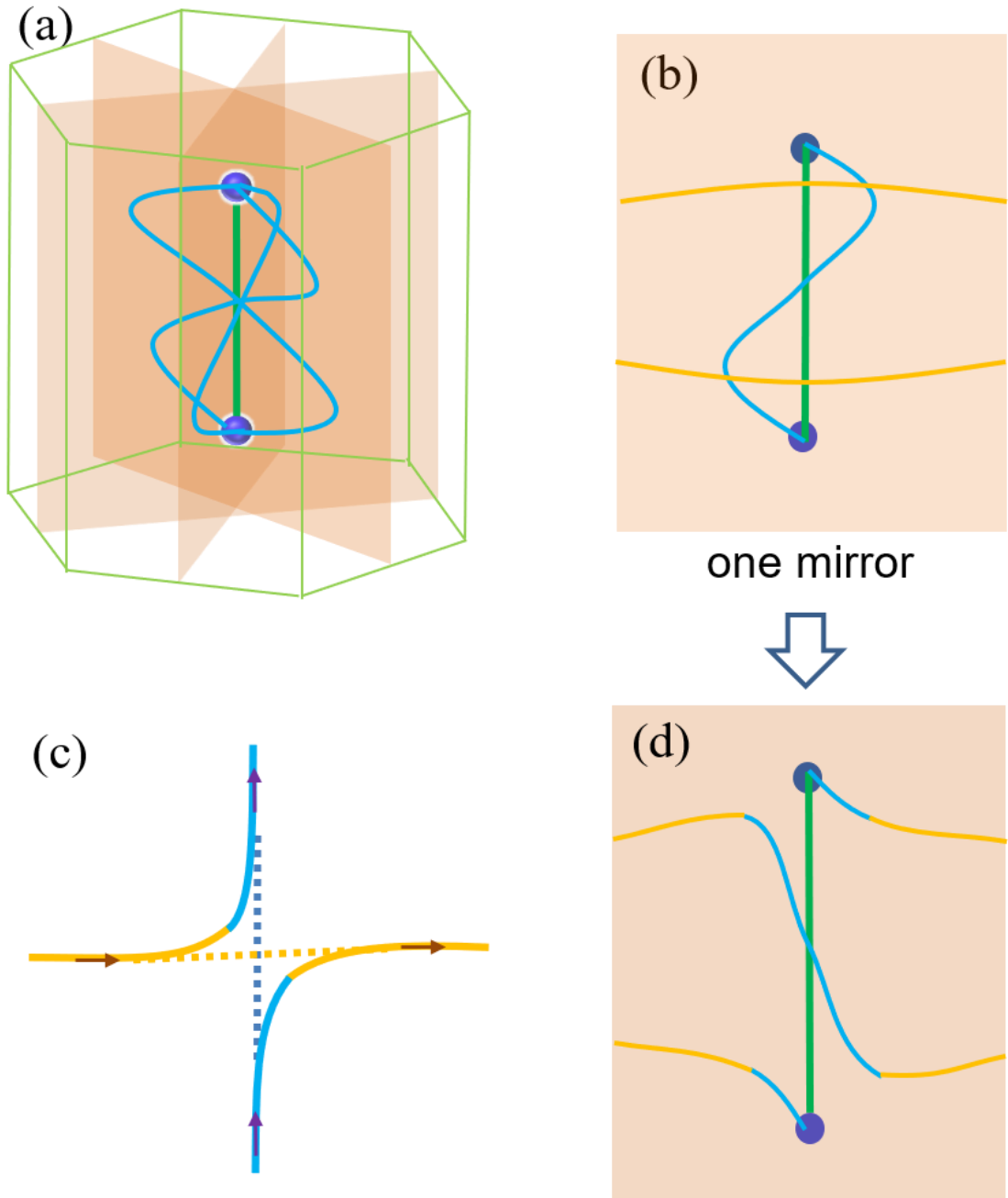

**Fig. 1. From standard NP phase to nexus network.** (a) A standard NP phase in which two NPs (blue dots) are connected by four nodal lines. The three light blue lines reside on the three mirror planes, respectively, while the green nodal line is shared by the three mirror planes. (b) A mirror plane of nexus phase plus two yellow topological ANLs. (c) Avoiding crossing of two (light blue and orange) topological nodal lines. (d) A mirror plane of nexus network in which the NPs are connected by anticrossing (winding) nodal lines.

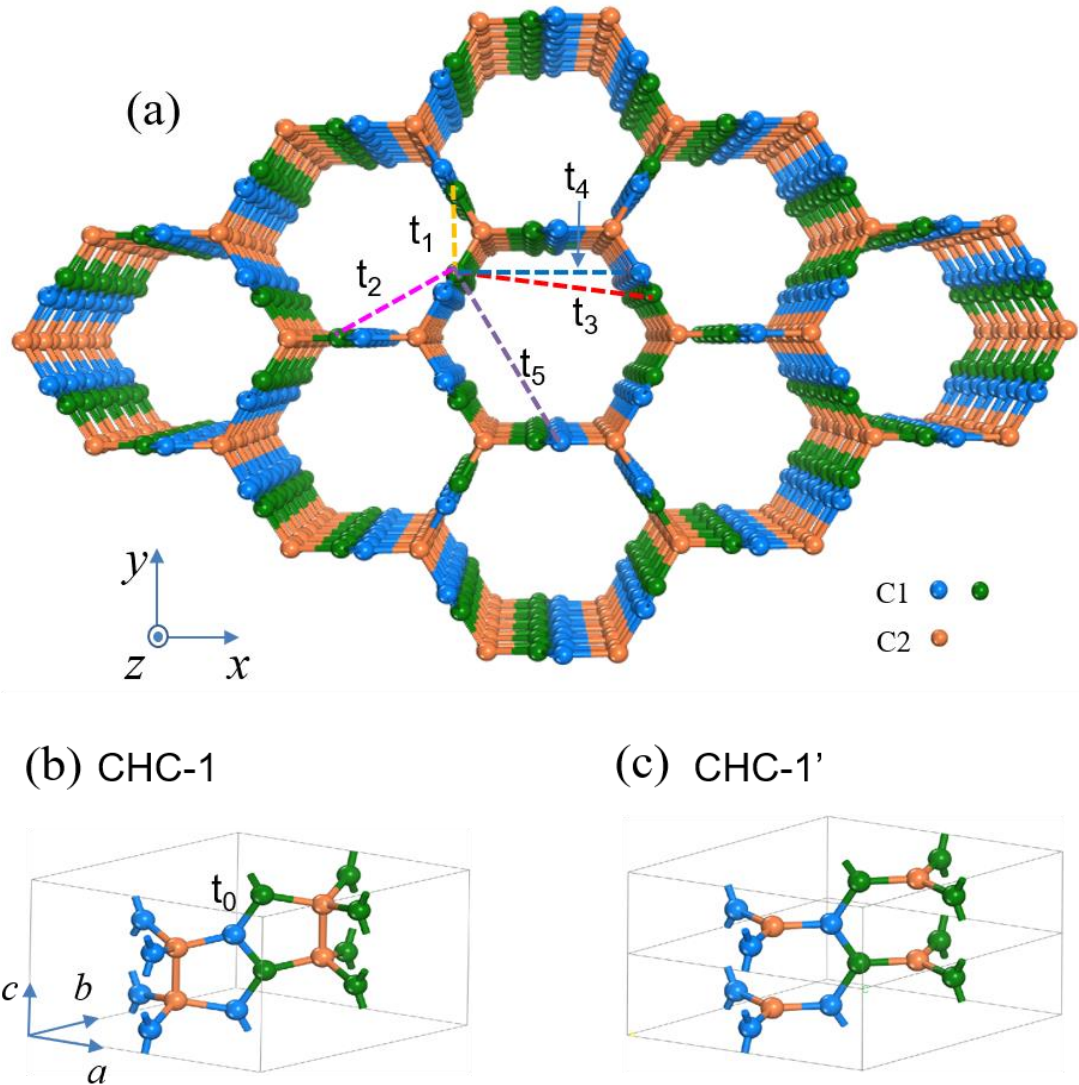

**Fig. 2. Atomic structures of CHCs.** (a) CHC-1 in a top view perspective, where the carbon atoms form a 3D honeycomb. The (green and blue) $sp^2$-carbon atoms (C1) reside on different horizontal atomic planes with respect to the c axis shown in panel (b) and each kind has a 3-fold rotational symmetry with respect to the axes passing through the (orange) $sp^3$-carbon atoms (C2). (b) The primitive cell for CHC-1, where *a*, *b*, and *c* are the lattice parameters. The C2 atoms form a row of carbon dimers along the *c* axis. The labels $t_0$ in (b) and $t_1$ to $t_5$ in (a) are tight-binding hopping parameters. (c) A (1×1×2) supercell of CHC-1′ which is a carbon structure after the dimers in CHC-1 are removed.

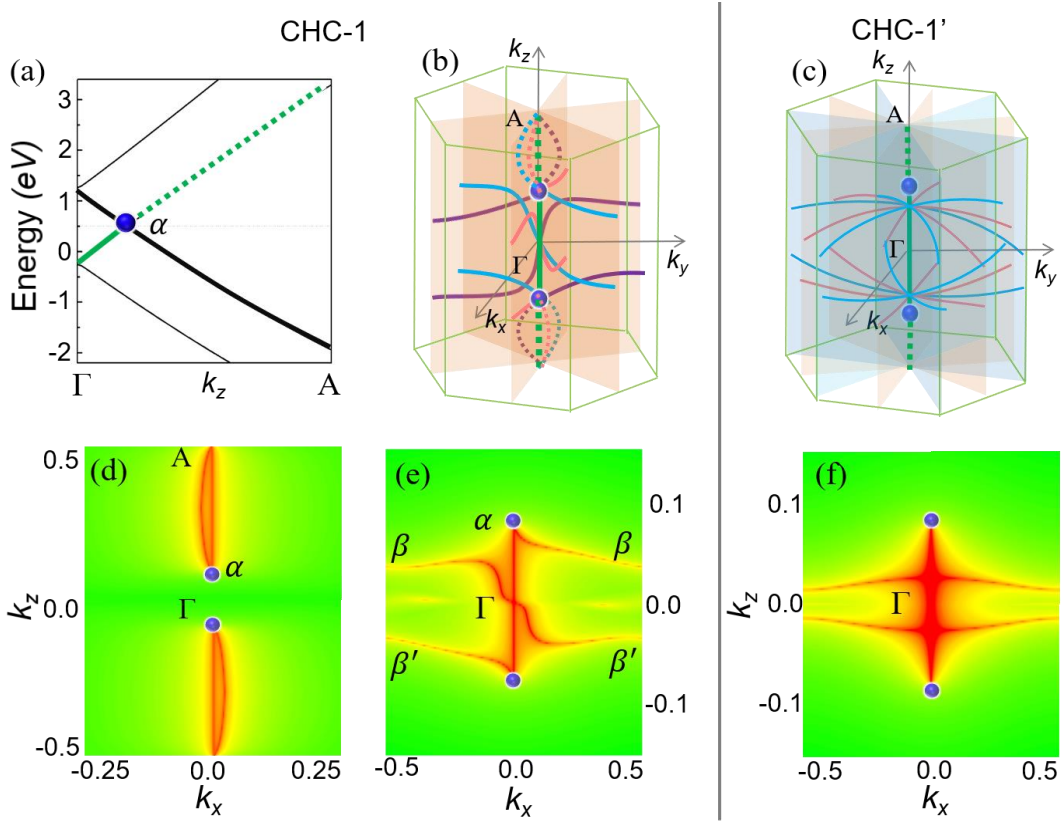

**Fig. 3. Band structure of CHC-1, and topological networks in CHC-1 and CHC-1'.** (a) Band structure of CHC-1 along $k_z$ (Γ − A). Point α (blue dot) is a TP, as a result of band crossing between the double degenerate (green) and non-degenerate (black) bands. (b) Schematic view of the nexus network including NPs and nodal lines in the full first BZ. The solid and dotted lines correspond to nodal lines for bands (*m-1,m*) and (*m,m+1*) (*m*=33), respectively. The three orange planes are mirror planes. (c) Schematic view of the TP-ANL in the full first BZ of CHC-1′. The three orange planes are mirror planes, while the three light blue planes are glide planes. In each plane, there are two nodal lines. (d) Contour plots of energy difference between bands m, m+1 on the mirror plane $k_y = 0$ for the structure CHC-1, which shows a standard connectivity between two NPs. The blue dots correspond to the NPs, while the red lines represent the energy difference is equal to zero, that is, nodal lines. (e) Same to (d) between bands m-1, m, which shows a winding connectivity between two NPs. (f) Contour plots of energy difference between bands m-1, m on the mirror plane $k_y = 0$ for the structure CHC-1′, where two TPs and two ANLs coexist. It is noted that, to clearly show the nodal lines in (d-f), different scales of $k_x$ and $k_z$ are applied. $k_x$ and $k_z$ in (d-f) are in units of $\pi/a$ and $\pi/c$ (*a* and *c* are lattice constants), respectively.

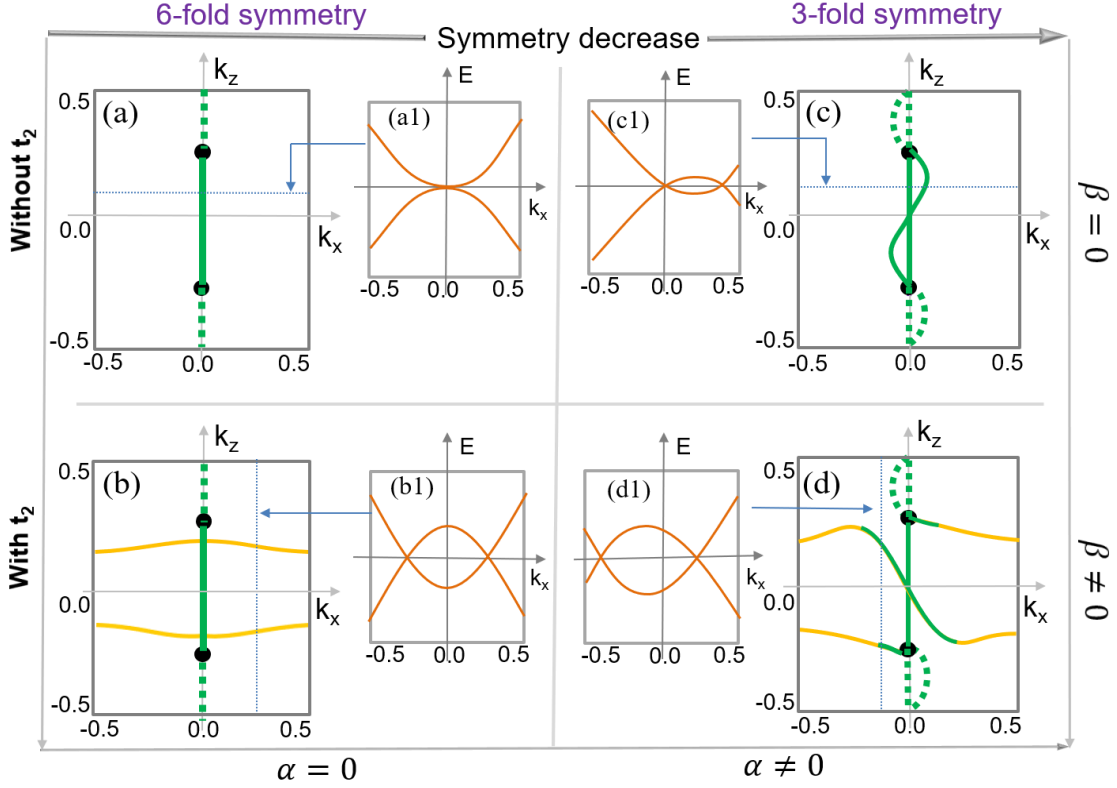

**Fig. 4. Four topological phases of the tight-binding model in Eq. (1) and k · p model in Eq. (2).** (a) A standard TP phase in which two TPs are connected by one trivial nodal line. (b) A TP-ANL phase in which TPs and (orange) ANLs coexist. (c) A standard NP phase in which NPs are connected by two splitting lines. (d) A nexus network in which NPs are connected by standard and winding connectivity. (a1-d1) (in the middle) Band structures corresponding to the k paths in (a-d), respectively, referred by the arrows. The standard TP phase in (a) and TP-ANL phase in (b) exist in the structure CHC-1′ with 6-fold screw rotational symmetry, while the NP phase in (c) and nexus network in (d) exist in the structure CHC-1 with 3-fold rotational symmetry. The phases are only shown on the $k_y = 0$ mirror plane, but identical features appear on the other five mirror planes for the phases in (a-b), and the other two mirror planes for those in (c-d). Different parameters for the tight-binding model and k · p model lead to the phase transitions between (a-d). In the tight-binding model in Eq. (1), $t_2$ determines if a phase has ANL or not, while the structural symmetry determines whether a phase is TP or nexus. In the k · p model in Eq. (2), in contrast, these different phases are determined by the vanishing or non-vanishing of β and α, respectively. $k_x$ and $k_z$ are in units of $\pi/a$ and $\pi/c$, respectively.

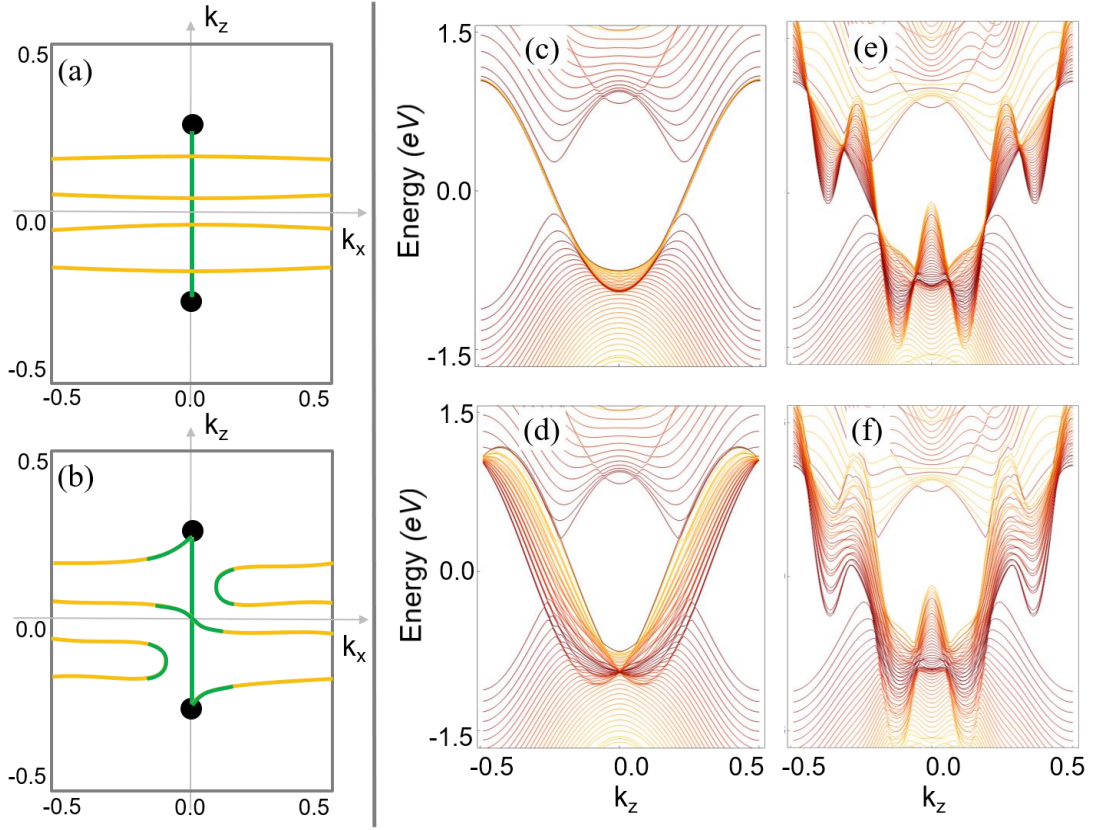

**Fig. 5. Extended TP-ANL phase and nexus network, and LLs based on the k · p model in Eq. (2) ($\beta$ is replaced by $\beta + \gamma\cos(\delta k_z)$).** (a) A TP-ANL phase in which a TP phase and four ANLs coexist ($\alpha = 0, \gamma = 0.5, \delta = 4$). (b) A nexus network evolved from (a) ($\alpha = 0.1, \gamma = 0.5, \delta = 4$). (c) LLs for the TP-ANL phase in Fig. 4(b) ($\alpha = 0$). (d) LLs for the nexus network in Fig. 4(d) ($\alpha = 0.5$), (e-f) LLs for the TP-ANL phase in Fig. 5(a) and the nexus network in Fig. 5(b), respectively. The other parameters are $A_1 = 1$, $B_1 = -1$, $A_2 = -1$, $B_2 = 1$, $D = 1$, $C_1 = C_2 = 0$, $\beta = -1$, and the magnetic field is B = 0.03. $k_x$ and $k_z$ are in units of $\pi/a$ and $\pi/c$, respectively.

# Supplementary Information

## I. Tight-binding parameters for CHC-1 (CHC-1')

The sp$^3$ dimer atoms in CHC-1 decrease the symmetry of the structure, which complicates the bond lengths between atoms. The primitive cell of CHC-1 has 12 sp$^2$ atoms, and these atoms can be divided into 4 layers along $c$ axis according to the colors. The atoms in the primitive cell form a triangle in each layer, as shown in Figs. S2(a-b). The structure has an inversion symmetry, and thus the first (second) layer is equal to the fourth (third) layer. However, the first layer is different from the second layer. The intra- and inter-layer bond lengths also depend on the layers. Therefore, there are multiple hopping energies between atoms. To describe the atomic interactions by the hopping parameters, based on $t_i$ (i=0~5), we define $t_{i/mn}$ (i=0~5, m, n=1~4), where m and n label the ordinal number of layers. For instance, in Figs. S2(a) and S2(b), we present two examples for $t_0$ and $t_1$. All the hopping parameters vary with the bond lengths. In Table S2, the corresponding hopping energies for CHC-1 and CHC-1' are shown. Figure S2(c) shows the band structure of CHC-1 according to the values given in the last column of Table S2. One can find that these parameters reproduce the DFT band structure in Fig. S1(a).

## II.  k · p model for CHC-1 (CHC-1')

We assume a three-band model expanded in the vicinity of the $k_z$ axis for a system having time-reversal (complex conjugation $K$), inversion ($I$), mirror ($M_y$), and 3-fold rotation ($C_3^z$) symmetries. The three basis states will be denoted as $|1\rangle = |v_-\rangle$, $|2\rangle = |v_+\rangle$, and $|3\rangle = |v_0\rangle$. For a heuristic picture, we can imagine that these have the same symmetries as $|p_y\rangle + i|p_x\rangle$, $|p_y\rangle - i|p_x\rangle$, and $|s\rangle$ respectively, where $|p_x\rangle$ and $|p_y\rangle$ refer to $p$ orbitals and $|s\rangle$ is an $s$ orbital, although our actual basis orbitals are more complicated linear combinations of atomic orbitals in the layered structure. The $3 \times 3$ matrix $H_{ij}(\boldsymbol{k})$ is Hermitian for any $\boldsymbol{k}$, i.e., $H_{ji}(\boldsymbol{k}) = H_{ij}^*(\boldsymbol{k})$.

Regarding inversion symmetry $I$, the orbitals are assumed to transform as $I|v_-\rangle = -|v_-\rangle$, $I|v_+\rangle = -|v_+\rangle$, and $I|v_0\rangle = |v_0\rangle$. Inversion takes $\mathbf{k} \to -\mathbf{k}$ in addition, so it follows that $H_{ij}(\boldsymbol{k}) = H_{ij}(-\boldsymbol{k})$ for $ij$=11, 22, 33, and 12, while $H_{ij}(\boldsymbol{k}) = -H_{ij}(-\boldsymbol{k})$ for $ij$=13 and 23.

Regarding time reversal symmetry K, the orbitals are assumed to transform as $K|v_-\rangle = |v_+\rangle$, $K|v_+\rangle = |v_-\rangle$, and $K|v_0\rangle = |v_0\rangle$. TR also takes $\mathbf{k} \to -\mathbf{k}$, and is antiunitary, so it follow that $H_{11}(\boldsymbol{k}) = H_{22}(-\boldsymbol{k})$, $H_{33}(\boldsymbol{k}) = H_{33}(-\boldsymbol{k})$, $H_{12}(\boldsymbol{k}) = H_{12}(-\boldsymbol{k})$, and $H_{13}(\boldsymbol{k}) = H_{23}^*(\boldsymbol{k})$.

Regarding rotational symmetry $C_3^z$, the orbitals transform are assumed to transform as $C_3^z|v_-\rangle = \gamma|v_-\rangle$, $C_3^z|v_+\rangle = \gamma^2|v_+\rangle$, and $C_3^z|v_0\rangle = |v_0\rangle$, where $\gamma = e^{2\pi i/3}$. $C_3^z$ also takes $k_- \to \gamma k_-$, $k_+ \to \gamma^2 k_+$, and $k_0 \to k_0$. Thus, if an element in the Harmiltonian matrix has a term of the form $k_-^{m_-} k_+^{m_+}$ then $m_+ - m_- = 2 \mod 3$ for $H_{13}$.

With all of these constraints, we can systematically write the Hamiltonian matrix up to overall quadratic order in $k_z$ and $k_y$ as

$$H(k) = \begin{pmatrix} a(k_z) + b(k_z)k_\parallel^2 & e'(k_z)k_+ + f(k_z)k_-^2 & g(k_z)k_- + h'(k_z)k_+^2 \\ e'(k_z)k_- + f(k_z)k_+^2 & a(k_z) + b(k_z)k_\parallel^2 & -g(k_z)k_+ + h'(k_z)k_-^2 \\ g(k_z)k_+ + h'(k_z)k_-^2 & g(k_z)k_- + h'(k_z)k_+^2 & c(k_z) + d(k_z)k_\parallel^2 \end{pmatrix}$$

where $k_\parallel^2 = k_+ k_- = k_x^2 + k_y^2$, and real $a(k_z)$ etc. are even functions of $k_z$ if they do not carry a prime and odd if they do.

Finally, we construct a minimal model by letting the odd functions of $k_z$ take the form $\sin k_z$ and even ones being $\cos k_z$ plus a constant. Also dropping the quadratic tem involving $h'(k_z)$ and renaming the coefficients, we obtain the model

$$H(k) = \begin{pmatrix} A_1 k_\parallel^2 + B_1 \cos k_z + C_1 & \alpha k_+ \sin k_z + \beta k_-^2 & Dk_- \\ \alpha k_- \sin k_z + \beta k_+^2 & A_1 k_\parallel^2 + B_1 \cos k_z + C_1 & -Dk_+ \\ Dk_+ & -Dk_- & A_2 k_\parallel^2 + B_2 \cos k_z + C_2 \end{pmatrix}, \tag{1}$$

that was introduced earlier in the main text.

Equation (1) can be simplified to the following form,

$$H(k) = \begin{pmatrix} Ak_\parallel^2 + B \cos k_z & \alpha k_+ \sin k_z + \beta k_-^2 & Dk_- \\ \alpha k_- \sin k_z + \beta k_+^2 & Ak_\parallel^2 + B \cos k_z & -Dk_+ \\ Dk_+ & -Dk_- & -Ak_\parallel^2 - B \cos k_z \end{pmatrix}. \tag{2}$$

By solving Eq. (2), one can obtain the three analytic energy eigenvalues $E_1$, $E_2$ and $E_3$. Then, the nodal line should satisfy $E_1 = E_2$, i.e.,

$$(\beta k_x^2 + \alpha k_x \sin k_z)[(2(Ak_x^2 + B \cos k_z) + (\beta k_x^2 + \alpha k_x \sin k_z)] = D^2 k_x^2. \tag{3}$$

It has four solutions, one of which is $k_x = 0$, corresponding to the straight nodal line along $k_z$. The other three correspond to the hybrid nodal lines on the plane $k_y = 0$. By tuning the parameters $\alpha$ and $\beta$, transitions among the four phases will result.

Based on Eq. (3), we can get the following information when A=D=1 and B=-1:

(1) When $\alpha = 0$, i.e., at the phase of triple points + nodal lines, the locations of the extra nodal lines are $\cos k_z = D/2B\beta$. Therefore, the extra nodal line shifts to high $k_z$ when $\beta$ is increased.

(2) At $\Gamma$, $k_z = (\frac{D}{2\alpha B} - \frac{\beta}{\alpha})k_x$, because $k_x \to 0, k_z \to 0$. Therefore, at the nexus phase, the slope of the nodal line is $\frac{D}{2\alpha B}$ because $\beta = 0$. At the nexus network phase, the slope changes its sign with the variation of $\beta$.

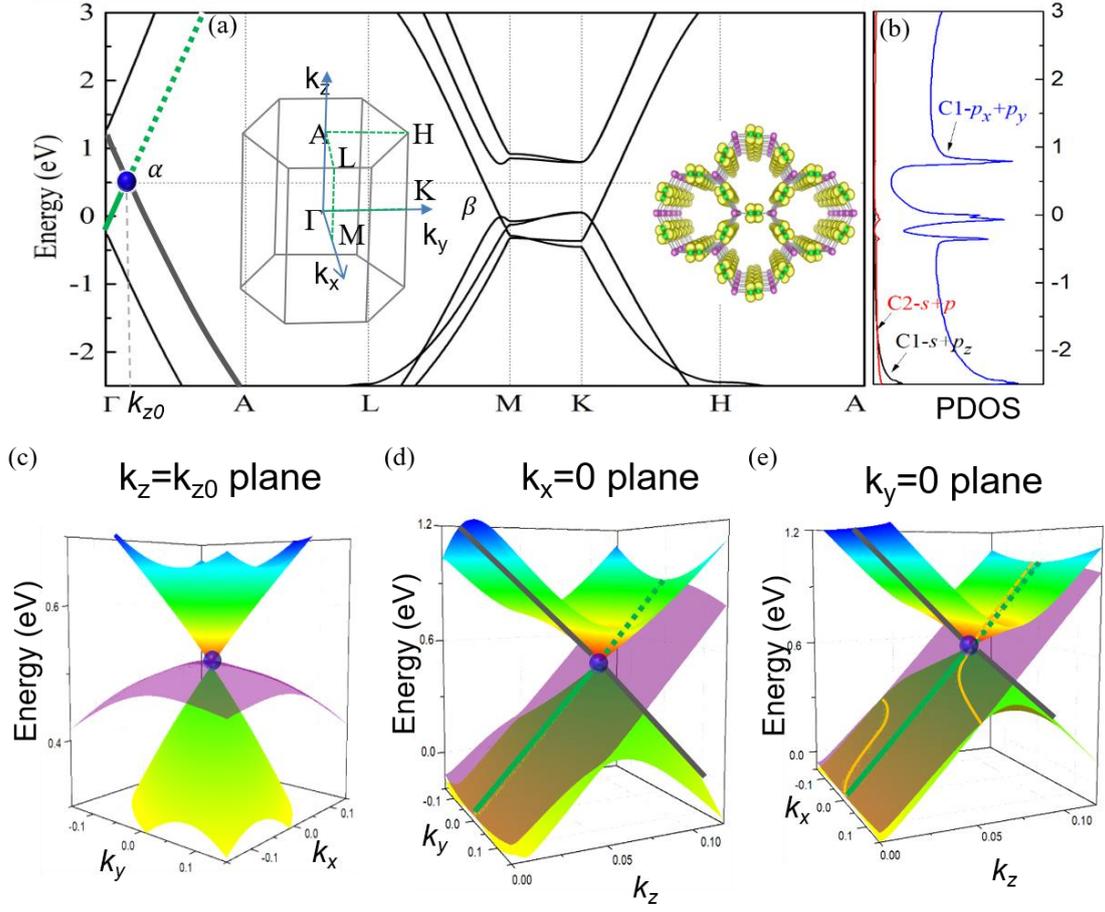

Fig. S1. (a) Band structure of CHC-1. Point α (blue dot at $k_z = 0.07$ $\pi/c$) along $\Gamma - A$ is a NP, as a result of band crossing between the double degenerate (green) and non-degenerate (red) bands. Inset (left): first BZ and inset (right): charge density contour of a state near α. (b) Partial density of states (PDOS) of CHC-1. (c)-(e) 3D band structures around α with (c) $k_z = 0.07$ $\pi/c$, (d) $k_x = 0$, and (e) $k_y = 0$, respectively. The green lines in (d) and (e) are where the tilted (purple) energy surface and the cones intercept. The color and line pattern are the same as in (a). The orange lines in (e) are where additional crossing lines between the energy surface and cones have been registered.

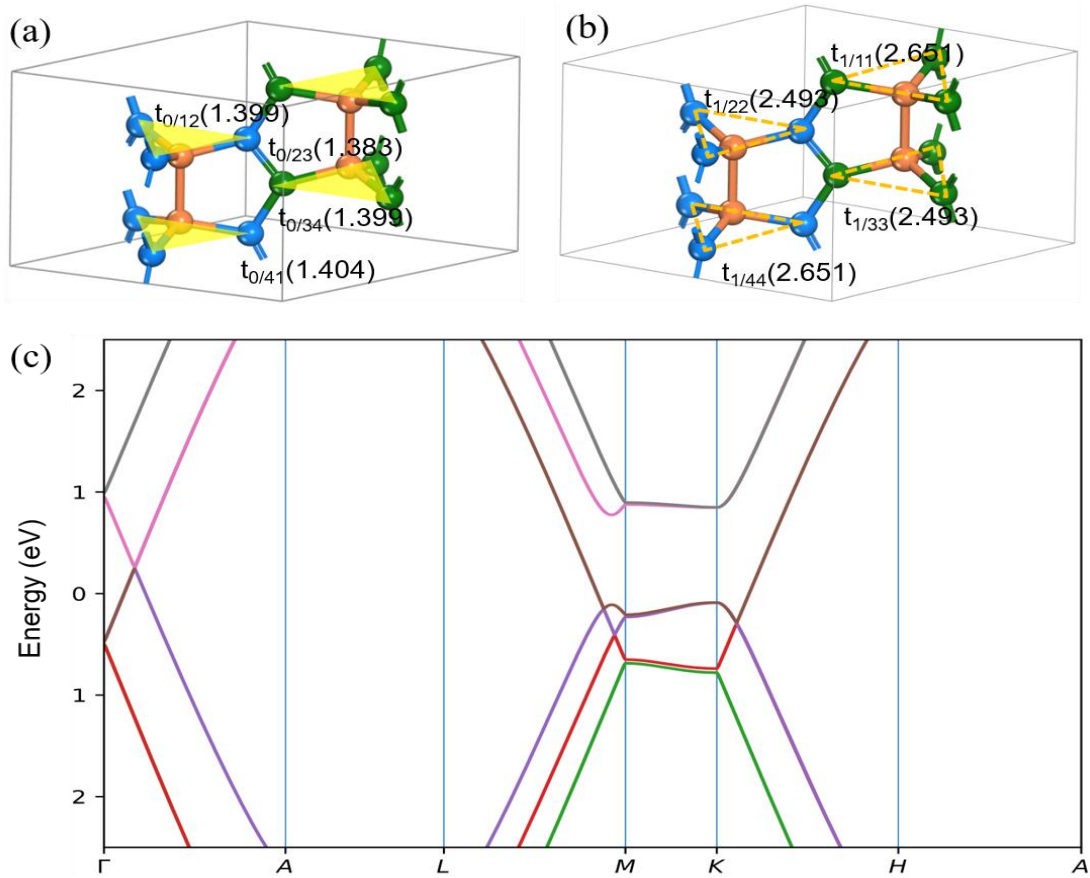

Fig. S2. (a-b) Two example for the tight-binding parameters $t_0$ and $t_1$ in the CHC-1. $t_{i/mn}$ (i=0~1, m, n=1~4) is the hopping energy $t_i$ between layers m and n. The yellow triangles in (a) and dashed orange triangles in (b) correspond to four atomic layers in the primitive cell of CHC-1. The values in the parentheses are corresponding bond lengths. (c) Band structure of CHC-1 by using the tight-binding parameters in the last column of Table S2.

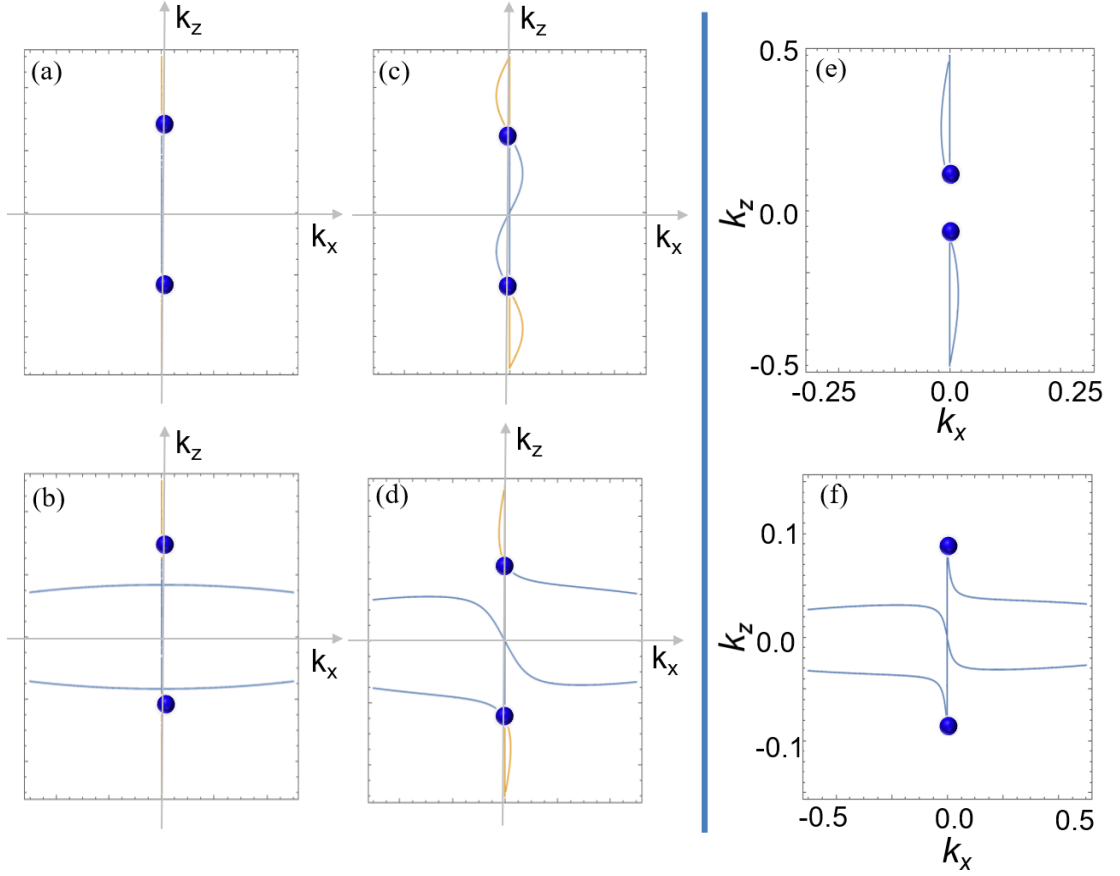

Fig. S3. (a-d) Four topological phases, based on the k·p model in Eq. (2), corresponding to the phases in Figs. 4(a-d), respectively. (a) A TP phase with $\alpha = \beta = 0$; (b) A TP-ANL phase with $\alpha = 0, \beta = -1$; (c) A NP phase with $\alpha = 0.2, \beta = 0$; (d) A nexus network with $\alpha = 0.2, \beta = -1$. The other parameters are $A_1 = 1$, $B_1 = -1$, $A_2 = -1$, $B_2 = 1$, $D = 1$, $C_1 = C_2 = 0$. (e-f) The nexus network similar to those in Figs. 3(d) and 3(e), reproduced by Eq. (2) with $A_1 = -A_2 = 1.45$, $B_1 = -B_2 = -1$, $D = 1$, $C_1 = -C_2 = 0.85$, $\alpha = 0.05$.

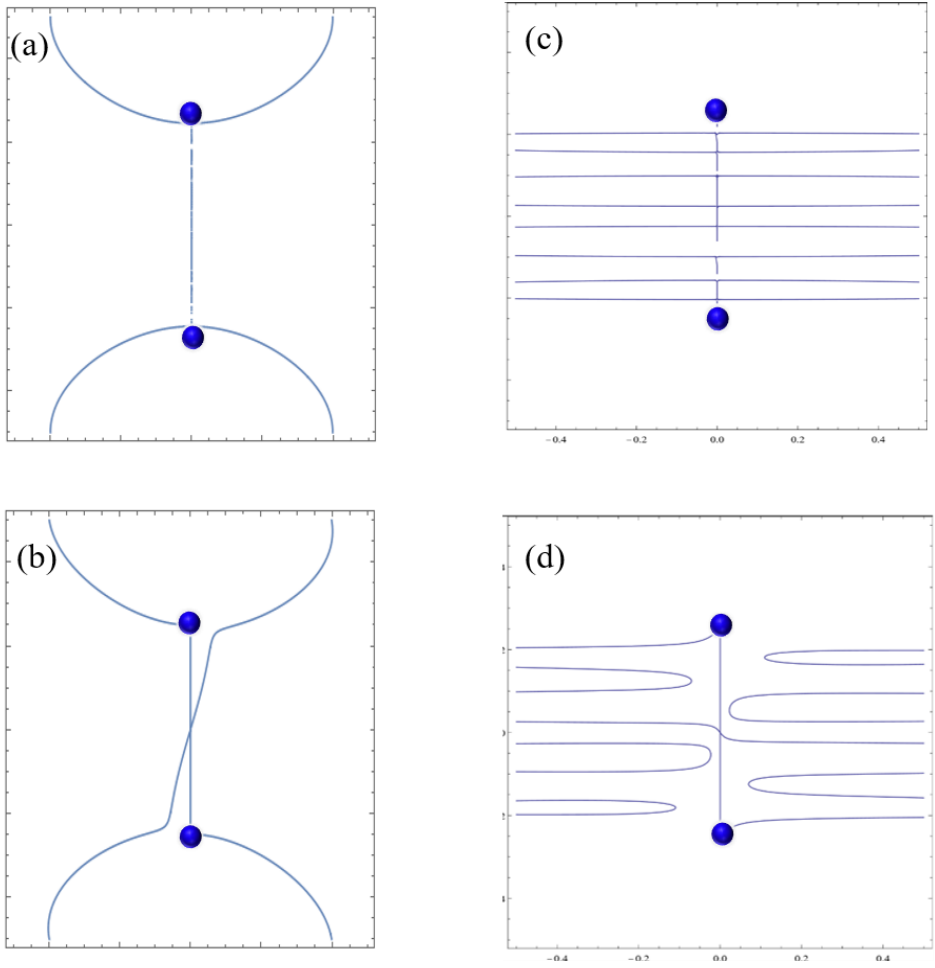

Fig. S4. Other TP-ANL and nexus network based on the $k \cdot p$ model in Eq. (2). (a) TP-ANL phase including two TPs and a nodal ring ($\alpha = 0$, $\beta = -15$); (b) Nexus network evolving from (a) when the TPs transition to NPs ($\alpha = 0.8$, $\beta = -15$); (c) TP-ANL phase including two TPs and additional 8 nodal lines ($\alpha = 0$, $\beta = -1$, $\gamma = 0.5$, $\delta = 12$); (d) Nexus network evolved from (c) when the TPs transition to NPs ($\alpha = 0.1$, $\beta = -1$, $\gamma = 0.5$, $\delta = 12$). For (c) and (d), $\beta$ in Eq. (2) is replaced by $\beta + \gamma \cos(\delta k_z)$. The other parameters are $A_1 = 1$, $B_1 = -1$, $A_2 = -1$, $B_2 = 1$, $D = 1$, $C_1 = C_2 = 0$.

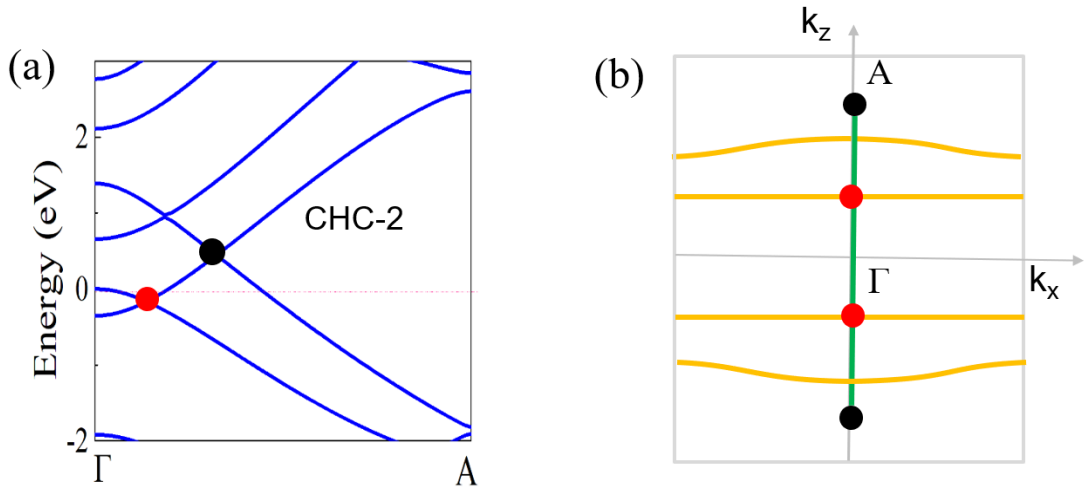

Fig. S5. (a) Band structure along $k_z$ (Γ − A) for CHC-2, in which there exist a TP (black dot) and a 4-fold degenerate point (red dot). (b) Topological phases of CHC-2 on the plane $k_y$=0. It is a TP-ANL phase including a TP, a 4-fold degenerate point and additional 4 nodal lines. The TP and 4-fold degenerate point are connected by a trivial (green) nodal lines.

Table S1. Structural parameters of CHC-1, CHC-2 and CHC-1'. The corresponding parameters of CKL, diamond and graphite are also shown for comparison.

| Structure | Space group | Lattice parameters | | Density | Bond lengths | Bulk Moduli | $E_{coh}$ |
|---|---|---|---|---|---|---|---|
| | | a = b | c | | | | |
| CHC-1 | P-3m1 | 6.35 | 4.83 | 1.89 | 1.38 ~ 1.64 | 225.00 | -7.57 |
| CHC-1' | P6$_3$/mmc | 6.39 | 2.42 | 1.87 | 1.41, 1.49 | 220.77 | -7.37 |
| CHC-2 | P6$_3$/mmc | 10.10 | 4.86 | 1.30 | 1.39 ~ 1.64 | 155.54 | -7.66 |
| CKL | P6$_3$/mmc | 4.46 | 2.53 | 2.75 | 1.50, 1.53 | 322 | -7.44 |
| Diamond | Fd$\bar{3}$m | 3.56 | 3.56 | 3.55 | 1.54 | 431.32 | -7.77 |
| Graphite | P6$_3$/mmc | 2.46 | 6.80 | 2.24 | 1.42 | 36.40 | -7.90 |

Table S2. Tight-binding hopping parameters $t_{i/mn}$ (i=0~5, m, n=1~4) in Eq. (1) for the four phases in Fig. 4, where m and n label the ordinal number of layers (see the explanation in Fig. S2). Parameters in the TP and TP-ANL phases are for the CHC-1' structure, while those in the NP and Nexus network phases are for the CHC-1 structure.

| $t_i$ | mn | TP phase | NP phase | TP-ANL phase | Nexus network |
|---|---|---|---|---|---|
| $t_0$ | 12 or 34 | 2.8 | 2.8 | 2.8 | 2.8 |
| | 23 | | 3.2 | | 3.2 |
| | 14 | | 2.4 | | 2.4 |
| $t_1$ | 11 or 44 | 0.45 | 0.4 | 0.45 | 0.4 |
| | 22 or 33 | | 0.5 | | 0.5 |
| $t_2$ | 11 or 44 | 0 | 0 | 0.15 | 0.18 |
| | 22 or 33 | | | | 0.12 |
| $t_3$ | 11 or 44 | -0.045 | -0.05 | -0.045 | -0.05 |
| | 22 or 33 | | -0.04 | | -0.04 |
| $t_4$ | 12 or 34 | -0.015 | -0.018 | -0.015 | -0.018 |
| | 23 | | -0.013 | | -0.013 |
| | 14 | | -0.013 | | -0.013 |
| $t_5$ | 12 or 34 | -0.02 | -0.025 | -0.02 | -0.025 |
| | 23 | | -0.013 | | -0.013 |
| | 14 | | -0.013 | | -0.013 |

Table S3. $\mathbf{k} \cdot \mathbf{p}$ model parameters in Eq. (2) for different topological phases in Figs. 4 and 5.

|  | TP phase in Fig. 4(a) | TP-ANL phase in Fig. 4(b) | NP phase in Fig. 4(c) | nexus network in Fig. 4(d) | Extended TP-ANL phase in Fig. 5(a) | Extended nexus network in Fig. 5(b) |
| --- | --- | --- | --- | --- | --- | --- |
| $A_1$ | 1 | 1 | 1 | 1 | 1 | 1 |
| $B_1$ | -1 | -1 | -1 | -1 | -1 | -1 |
| $C_1$ | 0 | 0 | 0 | 0 | 0 | 0 |
| $A_2$ | -1 | -1 | -1 | -1 | -1 | -1 |
| $B_2$ | 1 | 1 | 1 | 1 | 1 | 1 |
| $C_2$ | 0 | 0 | 0 | 0 | 0 | 0 |
| $D$ | 1 | 1 | 1 | 1 | 1 | 1 |
| $\alpha$ | 0 | 0 | 0.5 | 0.5 | 0 | 0.1 |
| $\beta$ | 0 | -1 | 0 | -1 | -1 | -1 |
| $\gamma$ | 0 | 0 | 0 | 0 | 0.5 | 0.5 |
| $\delta$ | 0 | 0 | 0 | 0 | 4 | 4 |